\documentclass[aps,prl,twocolumn,amsmath,amssymb,amsfonts,nofootinbib,long,floatfix]{revtex4}
\usepackage{epsfig,latexsym,bm}

\newcommand\kk{\mathbf{k}}

\begin{document}

\title{Reply to ``Comments on Origin of cosmic magnetic fields''}

\author{Leonardo Campanelli$^{1}$}
\email{leonardo.campanelli@ba.infn.it}
\affiliation{$^1$Dipartimento di Fisica, Universit\`{a} di Bari, I-70126 Bari, Italy}

\date{\today}

\maketitle


In~\cite{Campanelli}, we have shown that
cosmic magnetic fields are a natural consequence of inflation. Our results
have been criticized in a recent comment~\cite{Durrer}.
We show that the arguments raised against the validity of our results are incorrect.

1. 
The statement in~\cite{Durrer}, that at the beginning of
the inflationary expansion the modes that
are outside the (Hubble) horizon are not physically relevant,
is physically incorrect.
From the beginning to the end of inflation,
both sub- and superhorizon modes are physical.
This is because the presence of a Hubble horizon during an inflationary phase
(not necessarily a de Sitter phase)
just prevents two given observers separated at a certain time by a
physical distance greater than the Hubble radius $H^{-1}$
to ``communicate'' from that time on, to wit they are causally disconnected.
In particular, if they are causally disconnected at the beginning of inflation,
they will never communicate during inflation.
This, however, simply means that
superhorizon modes are not self-correlated,
in the sense that, even if they extend over regions of spacetime greater than the
Hubble volume, they cannot be used to causally connect regions in spacetime that are not
causally connected.
Moreover, inflation is not without end.
Modes that were superhorizon during inflation begin eventually to
re-enter the horizon after inflation, and early causally-disconnected observers
begin eventually to communicate.
Therefore, the introduction in~\cite{Durrer} of an infrared cut-off $k_{\rm min}$
to remove superhorizon modes at the beginning of inflation is physically
unjustified.

2. 
The magnetic power spectrum $\mathcal{P}_{\rm phys}(k,m)$ 
is not a physical quantity, 
as it is clearly seen from the fact that it depends on the
unphysical regulator photon mass $m$. It becomes a physical quantity
just in the limit of vanishing mass $m$, in which case it is positive defined.
Nevertheless, the procedure of renormalization 
could give rise to negative power spectra. 
There is, however, no physical and/or mathematical reason why a renormalized power spectrum
has to be positive defined.
In quantum physics in Minkowski spacetime the positivity of the power spectrum is assured
by the well-known Wiener-Khinchine theorem (see~\cite{Ford} for 
notations). When the applicability hypotheses of this theorem are satisfied,
we have a positive-defined spectrum, and this comes from the positivity
of the correlation function of Fourier transformed fields, $\mathcal{C}(t-t',\kk, \kk')$,
in the coincident limit $t \rightarrow t'$.
In quantum fields theory in curved spacetime, the applicability hypotheses of the Wiener-Khinchine theorem
are not generally satisfied in the case of renormalized correlators. This is because the renormalized
correlator $\mathcal{C}_{\rm phys}(0,\kk, \kk')$ is, in general, not positive defined in the limit $t \rightarrow t'$,
since it is given by difference between the exact correlation function and the adiabatic one,
$\mathcal{C}_{\rm phys}(0,\kk, \kk') = \langle \hat{F}^2(t,\kk) \rangle - \langle 0^{(A)}|(\hat{F}^{(A)})^2(t,\kk) |0^{(A)}\rangle^{(A)}$,
where $|0\rangle^{(A)}$ is the adiabatic vacuum and $\hat{F}$ some field operator.
This opens the possibility of having negative-defined power spectra.
According to the reasoning in~\cite{Durrer}, instead, one should expect
the vacuum expectation value (VEV) of a positive-defined quadratic quantity
like the energy density to be always positive.
The most famous example that invalidates this reasoning is the Casimir effect~\cite{Birrell-Davies}.
The electromagnetic energy density between two infinite parallel conducting plates
is positive defined but its VEV is ultraviolet divergent. After renormalization,
it becomes finite, negative, constant over space, and its effect (an attractive
force between the plates) has been confirmed experimentally.

3. 
The validity of the adiabatic renormalization procedure was criticized
in~\cite{Durrer1,Durrer2}. 
As stressed in~\cite{Parker et al},
the adiabatic renormalization is a {\it formal} procedure~\cite{Parker et al}: One subtracts
from the exact solution and mode-by-mode the corresponding approximate adiabatic solution,
the latter being generally a ``good'' approximation in the ultraviolet part of the spectrum.
However, in order to assure conservation of the renormalized stress tensor, one must
apply this subtraction also to modes in the infrared part of the spectrum.
When the adiabatic subtraction is applied, as postulated in~\cite{Durrer1,Durrer2}, only to
high-momentum modes for which the adiabatic approximation is mathematically accurate,
it amounts to the introduction of an effective (time-dependent) infrared cut-off
for the adiabatic counter-terms,
and it is then accompanied by the physically unacceptable result of having a non-conserved
renormalized stress tensor~\cite{Bastero-Gil}.

In conclusion, the criticisms in~\cite{Durrer}
are physically and mathematically unfounded.
Therefore, we stand by our original results and conclusions.


\end{document}